\documentclass[letterpaper,12pt]{article}
\usepackage{amssymb}
\usepackage{amsmath}
\usepackage[nolists,nomarkers]{endfloat}
\usepackage{fixmath}
\usepackage[letterpaper,nohead,nomarginpar,centering,margin=1in]{geometry}
\usepackage{graphicx}
\usepackage[numbers,square,sort&compress]{natbib}
\usepackage{setspace}

\DeclareMathOperator{\spn}{span}

\DeclareMathOperator{\diag}{diag}
\begin{document}

\title{Nonlinear Laplacian spectral analysis: Capturing intermittent and low-frequency  spatiotemporal patterns in high-dimensional data}

\author{Dimitrios Giannakis,\thanks{dimitris@cims.nyu.edu} \quad Andrew J.\ Majda}

\date{\emph{Courant Institute of Mathematical Sciences, New York University}\\
  \emph{251 Mercer St, New York, NY 10012}}

\maketitle

\begin{abstract}
 We present a technique for spatiotemporal data analysis called nonlinear Laplacian spectral analysis (NLSA), which generalizes singular spectrum analysis (SSA) to take into account the nonlinear manifold structure of complex datasets. The key principle underlying NLSA is that the functions used to represent temporal patterns should exhibit a degree of smoothness on the nonlinear data manifold $ M $; a constraint absent from classical SSA. NLSA enforces such a notion of smoothness by requiring that temporal patterns belong in low-dimensional Hilbert spaces $V_l$ spanned by the leading $l$ Laplace-Beltrami eigenfunctions on $ M $. These eigenfunctions can be evaluated efficiently in high ambient-space dimensions using sparse graph-theoretic algorithms. Moreover, they provide orthonormal bases to expand a family of linear maps, whose singular value decomposition leads to sets of spatiotemporal patterns at progressively finer resolution on the data manifold. The Riemannian measure of $ M $ and an adaptive graph kernel width enhances the capability of NLSA to detect important nonlinear processes, including intermittency and rare events. The minimum dimension of $V_l$  required to capture these features while avoiding overfitting is estimated here using spectral entropy criteria. 

As an application, we study the upper-ocean temperature in the North Pacific sector of a 700-year control run of the CCSM3 climate model. Besides the familiar annual and decadal modes, NLSA recovers a family of intermittent processes associated with the Kuroshio current and the subtropical and subpolar gyres. These processes carry little variance (and are therefore not captured by SSA), yet their dynamical role is expected to be significant.
\end{abstract}

\subsubsection*{Keywords}
Laplacian eigenmaps, singular spectrum analysis, intermittency, decadal variability, manifold embedding

\section{Introduction}

In recent years, there has been a proliferation of data in geophysics and other applied sciences acquired through observations \citep[][]{WDS11}, reanalyses \citep[][]{DeeEtAl11}, and large-scale numerical models \citep[][]{TaylorEtAl11}. The availability of such data holds promise for advances in a number of important applications, such as decadal-range climate forecasts through improved understanding of regime transitions in the ocean \citep[][]{TengBranstator10} and weather forecasts beyond seven-day lead times through skillful models of organized convection in the tropics \citep[][]{Roundy11}. However, a major obstacle in using large-scale datasets in such applications is their sheer volume and complexity. Typically, the data are presented in the form of a high-dimensional time series, acquired through noisy, partial observations of a strongly-nonlinear dynamical system. Thus, there exists a strong practical interest in developing novel data analysis techniques to decompose the data into a set of spatiotemporal patterns revealing the operating nonlinear processes. Such patterns can be used to gain scientific understanding of complex phenomena, or to build reduced dynamical models for prediction.      

Machine learning methods, such as those based on kernels \citep[][]{LimaEtAl09} and neural networks \citep[][]{Hsieh07}, are well-suited to capture the nonlinear features of data generated by complex systems, but in certain cases are prone to overfitting and/or poor scaling with the dimension of ambient space \citep[][]{Christiansen05}. In contrast, classical linear approaches, such as singular spectrum analysis (SSA) and its variants \cite{VautardGhil89,AubryEtAl91,GolyandinaEtAl01,GhilEtAl02}, have the advantages of direct physical interpretability and analysis through the tools of linear algebra, but are generally limited in detecting patterns carrying a high portion of the variance (energy) of the observed signal. Such patterns are likely to miss important nonlinear dynamical features, including intermittency and rare events \citep[][]{AubryEtAl93,CrommelinMajda04}. The latter carry low variance, but can play an important dynamical role by triggering large-scale regime transitions. 

In \citep[][]{GiannakisMajda12a} (building on preliminary work in \citep[][]{GiannakisMajda11}), a method called nonlinear Laplacian spectral analysis (NLSA) was developed, whose objective is to address the above shortcomings by combining aspects of both nonlinear and linear methods. Similarly to SSA, NLSA decomposes an observed signal through spectral analysis of linear operators mapping a space of temporal modes (the ``chronos'' space) to the corresponding space of spatial patterns (``topos'' space) \citep[][]{AubryEtAl91}. However, the linear operators used in NLSA, denoted here by $ A^l $, differ crucially from SSA in that they are tailored to the nonlinear manifold structure of the data. Specifically, the chronos spaces in NLSA are low-dimensional subspaces of the Hilbert space $ L^2( M, \mu ) $ of square-integrable functions on the data manifold $ M $, with $ \mu $ the Riemannian measure induced by the embedding of $ M $ in $ n $-dimensional ambient space. NLSA employs graph-theoretic algorithms \citep[][]{BelkinNiyogi03,CoifmanLafon06} to produce a set of orthonormal Laplace-Beltrami eigenfunctions, which form bases for a family of $ l $-dimensional subspaces $ V_l $  of  $ L^2( M, \mu ) $, with $ l $ controlling the scale (``resolution'') on the data manifold of the temporal patterns described by the corresponding $ A^l $ operator. A decomposition of the data into a set of  $ l $ spatial and temporal patterns then follows from singular value decomposition (SVD) of the $ n \times l $ matrix representing $ A^l $ in the eigenfunction basis. 

A key difference between NLSA and other nonlinear dimensionality reduction techniques in the literature is that the eigenfunctions are not used to define feature maps as done, e.g., in kernel PCA \citep[][]{ScholkopfEtAl98,LimaEtAl09}. Moreover, the diffusion process on the graph employed to evaluate the Laplace-Beltrami eigenfunctions has no implied relation to the actual dynamics (cf.\ \citep[][]{NadlerEtAl06,SingerEtAl09}). Rather, the graph Laplacian eigenfunctions are used in NLSA solely as a set of basis functions for $ V_l $.  The physical, spatiotemporal processes operating on the data manifold are obtained in the linear SVD step. Two further important ingredients of the scheme are: 
\begin{enumerate}
  \item Time-lagged embedding \citep[][]{SauerEtAl91} to alleviate non-Markovianity of the input data due to partial observations, and capture propagating coherent structures;
  \item Adaptive kernels in the construction of the graph Laplacian (with Gaussian widths determined from the distances between temporal nearest neighbors), enhancing the capability of the algorithm to recover rapid transitions and rare events.
\end{enumerate}

We demonstrate the efficacy of the scheme in an analysis of the North Pacific sector of the Community Climate System model version 3 (CCSM3) \cite{CollinsEtAl06}, augmenting the work in \citep[][]{GiannakisMajda12a,GiannakisMajda11} with new practical criteria for choosing the dimension $ l $ of the temporal spaces $ V_l $. Using a 700-year equilibrated dataset of the upper-300 m ocean \cite{AlexanderEtAl06,TengBranstator10,BranstatorTeng10}, we identify a number of qualitatively-distinct spatiotemporal processes, each with a meaningful physical interpretation. These include the seasonal cycle, semiannual variability, as well as decadal-scale processes resembling the Pacific decadal oscillation (PDO) \citep[][]{OverlandEtAl08}. Besides these modes, which are familiar from SSA, the spectrum of NLSA also contains modes with a strongly intermittent behavior in the temporal domain, characterized by five-year periods of high-amplitude oscillations with annual and semiannual frequencies, separated by periods of quiescence. Spatially, these modes describe enhanced eastward transport in the region of the Kuroshio current, as well as retrograde (westward) propagating temperature anomalies and circulation patterns resembling the subpolar and subtropical gyres. The bursting-like behavior of these modes, a hallmark of nonlinear dynamics, means that they carry little variance of the raw signal (about an order of magnitude less than the seasonal and PDO modes), and as a result, they are not captured by linear SSA. 

Here, we pay particular attention to the choice of the dimension  of $ V_l $. Introducing a spectral entropy measure $ D_l $ characterizing the change in the  energy distribution among the modes of $ A^l $ as $ l $ grows, we propose to select $ l $ as the minimum value beyond which $ D_l $ becomes small. We find this to be a particularly effective way to prevent overfitting the data (a common issue in machine learning methods \citep[][]{Christiansen05}), while capturing the important features of the signal through the spatiotemporal modes of $ A^l $.  Another key theme of the framework presented here is to apply minimal or no preprocessing to the data prior to analysis. In particular, we find that removing the seasonal cycle from the data (as is commonly done in the geosciences) eliminates the intrinsically nonlinear intermittent modes from the spectrum of $ A^l $; a fact which we attribute to the distortion of neighborhood relationships imparted by preprocessing a nonlinear data manifold through a linear operation such as seasonal detrending.  These results should be useful in several disciplines where data detrending is common practice. Overall, the theoretical perspective in this paper complements our earlier work \cite{GiannakisMajda12a}, emphasizing the role of regularity of the temporal basis functions on the nonlinear data manifold. 

The plan of this paper is as follows. In Sec.~\ref{secTheory}, we describe the theoretical framework of NLSA algorithms. In Sec.~\ref{secModes}, we apply this framework to the upper-ocean temperature in the North Pacific sector of CCSM3. We discuss the implications of these results in Sec.~\ref{secDiscussion}, and conclude in Sec.~\ref{secConclusions}. A Movie showing dynamical evolution of spatiotemporal patterns is provided as Additional Supporting Information.

\section{\label{secTheory}Theoretical framework}

We consider that we have at our disposal samples of a time-series $ x_t $ of a $ d $-dimensional variable sampled uniformly with time step $ \delta t $. Here, $ x_t \in \mathbb{ R }^d $ is generated by a dynamical system, but observations of $ x_t $ alone are not sufficient to uniquely determine the state of the system in phase space; i.e., our observations are incomplete. For instance, in Section~\ref{secModes} ahead, $ x_t $ will be  a depth-averaged ocean temperature field restricted in the North-Pacific sector of CCSM3. Our objective is to produce a decomposition of $ x_t $ into a set of $ l $ spatiotemporal patterns,
\begin{equation}
  \label{eqXTK}
  x_t \approx \sum_{k=1}^l \tilde x_t^k,
\end{equation}
 taking explicitly into account the fact that the underlying trajectory of the dynamical system lies on a nonlinear manifold $ M $ in phase space.   

\subsection{Overview of NLSA}

The methodology employed here to address this objective consists of four basic steps: (1) Embed the observed data in a vector space $ H $ of dimension greater than $ d $ via the method of delays; (2) construct a linear map $ A^l $ taking a Hilbert space of scalar functions on $ M $ representing temporal patterns to the spatial patterns in $ H $; (3) perform an SVD in a basis of orthonormal Laplacian eigenfunctions to extract the spatial and temporal modes associated with $ A^l $; (4) project the modes from $ H $ to physical space $ \mathbb{ R }^d $ to obtain the spatiotemporal patterns $ \tilde x_t^k $ in~\eqref{eqXTK}. Below, we provide a description of each step. Further details of the procedure, as well as pseudocode, are presented in \citep[][]{GiannakisMajda12a}.  Hereafter, we shall consider that $ M $ is compact and smooth, so that a well-defined spectral theory exists \cite{Berard86}.  Even though these conditions may not be fulfilled in practice, eventually we will pass to a discrete, graph-theoretic description \cite{Chung97}, where smoothness is not an issue. 

Step (1) is familiar from the qualitative theory of dynamical systems \cite{PackardEtAl80,Takens81,SauerEtAl91,DeyleSugihara11}. Under generic conditions, the image of $ x_t $ in embedding space $ H = \mathbb{ R }^n $ under the delayed-coordinate mapping,
\begin{equation}
  \label{eqEmbedding}
  x_t \mapsto X_t =
  \begin{bmatrix}
    x_t \\
    x_{t-\delta t } \\
    \vdots \\
    x_{t-(q-1) \, \delta t } 
  \end{bmatrix},
\end{equation}
 lies on a manifold which is diffeomorphic to $ M $ (i.e., indistinguishable from $ M $ from the point of view of differential geometry), provided that the dimension $ n $ of $ H $ is sufficiently large. Thus, given a sufficiently-long embedding window  $ \Delta t = ( q - 1 ) \, \delta t $, we obtain a representation of the nonlinear manifold underlying our incomplete observations, which can be thought of as a curved  hypersurface in Euclidean space. That hypersurface inherits a Riemannian metric $ g $ (i.e., an inner product between tangent vectors on $ M $ constructed from the canonical inner product of $ H $) and a corresponding Riemannian measure $ \mu = ( \det g )^{1/2} $.   

Steps~(2) and~(3) effectively constitute a generalization of SSA, adapted to nonlinear datasets. First, recall that SSA \citep[][]{AubryEtAl91}  views the data matrix 
\begin{equation}
  \label{eqX}
  X = [ X_0, X_{\delta t }, \ldots, X_{(s-1)\delta t} ] 
\end{equation}
(dimensioned $ n \times s $ for $ s $ samples in $ n $-dimensional embedding space) as a linear map from the space of temporal patterns $ \mathbb{ R }^s $ (so-called chronos space) to the space of spatial patterns $ H $ (the topos space), defined as
\begin{equation}
  \label{eqSSA}
  y =  X f.
\end{equation}
That is, the spatial pattern $ y \in \mathbb{ R }^n $ corresponding to the temporal pattern $ f = ( f_1, \ldots, f_s )^T \in \mathbb{ R }^s $ is given by a weighted sum of the data points $ X_{(i-1)\, \delta t} $ by $ f_i $. Depending on the application at hand, $ \mathbb{ R }^s $ may be replaced by a more general Hilbert space $ L^2( \mathcal{ T } ) $ over the set of temporal observations $ \mathcal{ T } = \{ 0, \delta t, \ldots, (s-1) \, \delta t \}  $. In either case, SSA produces a spatiotemporal decomposition of the signal through SVD of the linear map $ X $, viz. 
\begin{equation}
  \label{eqSVD}
  X = U \Sigma V^T,
\end{equation}
with
\begin{equation}
  \begin{gathered}
    U = [ u_1, \ldots, u_n ], \quad \Sigma = \diag( \sigma_1, \ldots, \sigma_{\min\{n,s\}} ), \quad  V = [ v_1, \ldots, v_s ], \\
    u_i \in H, \quad  \sigma_i \geq 0, \quad v_i \in L^2( \mathcal{ T } ).
  \end{gathered}
\end{equation} 
This leads to a rank-$ l $ decomposition of the signal through
\begin{equation}
  \label{eqXKTSSA}
  X^l_t = \sum_{k=1}^l \tilde X_t^k, \quad \tilde X_t^k = u_k \sigma_k v_k( t ),
\end{equation}
where $ v_k( t )$ is the component of $ v_k $ associated with time $ t \in \mathcal{ T } $. Similarly to~\eqref{eqSSA}, the spatial pattern $ u_k $ corresponding to $ v_k $ is given by a weighted sum of the input data, 
\begin{equation}
  \sigma_k u_k = X v_k.
\end{equation}
Geometrically, $ u_i $ are the principal axes of the covariance ellipsoid $ X X^T $, and $ \sigma v_k $ linear projections of the data onto those axes.

Even though the temporal patterns $ v_k $ are well-behaved functions of time [because they are square-integrable functions in $ L^2( \mathcal{ T } ) $], they exhibit no notion of regularity on the nonlinear data manifold. That is, the corresponding temporal patterns $ v_k( t ) $ in SSA (acquired through linear projections) may develop rapid oscillations  which are not related to the intrinsic geometry of the nonlinear data manifold (e..g, if the embedding of $ M $ in $ \mathbb{ R }^n $ produces folds). In NLSA, geometrical regularity is viewed as an essential ingredient of an efficient description of high-dimensional complex data, and is enforced by replacing the chronos space of SSA with function spaces on the data manifold of sufficient smoothness. More specifically, the temporal modes in NLSA have well-behaved derivatives on the data manifold, in the sense that $ \lVert v_k \rVert^2 =  \sum_a  \nabla^a v_k \, \nabla_a v_k $ is integrable on $ M $. Here and in~\eqref{eqLaplF},  summation over the tensorial indices $ a $ and $ b $ is over the dimensions of $ M $, and $ \nabla_a $ the gradient operator associated with the Riemannian metric $ g $ of $ M $. Moreover, we assume for now that $ M $ is smooth and compact, so that a well-defined spectral theory exists. Later, we will pass to a graph-theoretic description, where smoothness is not an issue. 

A natural set of basis functions possessing this type of regularity are  the eigenfunctions $ \phi_i $ of the Laplace-Beltrami operator $ \upDelta $, defined as \cite{Berard86}
\begin{equation}
  \label{eqLaplF}
 \upDelta( f ) = - \frac{ 1 }{ \mu } \sum_{a,b} \nabla_a \left( \mu g^{ab} \nabla_b f  \right), 
\end{equation}
with $ \sum_b g^{ab} g_{bc} = {\delta^a}_c $, and $ f $ an element of the Hilbert space $ L^2( M, \mu ) $ of square-integrable scalar functions on $ M $ with inner product inherited from the Riemannian measure $ \mu $ [see~\eqref{eqOrthonormality} ahead]. The eigenfunctions of $ \upDelta$ are solutions of the eigenvalue problem
\begin{equation}
  \label{eqLapl}
  \upDelta \phi_i = \lambda_i \phi_i 
\end{equation}
(together with appropriate boundary conditions if $ M $ has boundaries) with $ 0 = \lambda_0 < \lambda_1 \leq \lambda_2 \leq \cdots $. Moreover, $ \phi_i $ can be chosen to be orthonormal with respect to the inner product of $ L^2( M, \mu ) $, i.e.,   
\begin{equation}
  \label{eqOrthonormality}
  \int_M \mu( X )  \phi_i( X ) \phi_j( X ) = \delta_{ij}. 
\end{equation}

Let $ \Phi_l $ be the $ l $-dimensional subspace of $ L^2( M, \mu ) $ spanned by the leading $ l $ eigenfunctions from~\eqref{eqLapl} meeting the orthonormality condition in~\eqref{eqOrthonormality}; i.e., 
 \begin{equation}
   \label{eqPhiL}
   \Phi_l = \spn\{ \phi_0, \ldots, \phi_{l-1} \}.
\end{equation}
These spaces have the following important properties. 
\begin{enumerate}
\item As $ l \to \infty $, $ \Phi_l $ provides a dense coverage of $ L^2( M, \mu ) $. Moreover, if $ M $ is a differentiable manifold, every basis element of $ \Phi_l $ is a smooth function with bounded covariant derivative. Heuristically, $ l $ may be though of as a parameter controlling the scale on the data manifold resolved by the eigenfunctions spanning $ \Phi_l $.
\item For sufficiently small $ l $ and large-enough number of samples $ s $, the values of $ \{ \phi_0, \ldots, \phi_{l-1} \} $ on the discrete samples of the data manifold in~\eqref{eqX} can be computed efficiently using sparse graph-theoretic algorithms \citep[][]{BelkinNiyogi03,CoifmanEtAl05,CoifmanLafon06}. Even if $ M $ is not a smooth manifold (as is frequently the case in practice), the leading few eigenfunctions determined by graph-theoretic analysis can be associated with some smooth coarse-grained manifold reflecting the large-scale nonlinear geometry of $ M $.    
\end{enumerate}
Note that $ l $ is in general not equal to the dimension of $ M $.

In NLSA, the family of $ \Phi_l $ with $ l $ between one and some upper limit are employed to construct temporal spaces analogous to the chronos space in classical SSA. Specifically, given a trajectory $ X_t $ on the data manifold, a function 
\begin{equation}
  \label{eqF}
  f = \sum_{i=1}^l c_i \phi_i, \quad f \in \Phi_l
\end{equation}
 with expansion coefficients $ c_i $ gives rise to a temporal process $ \tilde f( t ) = f( X_t ) $ for $ t \in \mathcal{ T } $. Thus, introducing the Hilbert space $ L^2( \mathcal{ T }, \mu ) $ with weighted inner product
\begin{equation}
  \label{eqInnerProdVL}
  \langle \tilde f_1, \tilde f_2 \rangle = \sum_{t \in \mathcal{ T } } \mu( X_t ) \tilde f_1( t ) \tilde f_2( t ), 
\end{equation}
the $ l $-th temporal space in NLSA is the $ l $-dimensional subspace $ V_l $ of $ L^2( \mathcal{ T }, \mu ) $ consisting of temporal patterns $ \tilde f $ generated by functions of the form in~\eqref{eqF}.  

Similarly to~\eqref{eqSSA}, for each $ V_l $ there exists a linear map $ A^l : V_l \mapsto H $ linking the temporal modes to the spatial modes through the formula
\begin{equation}
  \label{eqAL}
  y = A^l( \tilde f ) = \sum_{t \in \mathcal{ T } } \mu( X_t ) X_t \tilde f( t ).
\end{equation}
Let $ \{ \phi_0, \ldots \phi_{l-1} \} $ be the orthonormal basis of $ \Phi_l $ from~\eqref{eqLapl}, and $ \{ e_1, \ldots, e_n \} $ an orthonormal basis of $ H $. The matrix elements of $ A^l $ with this choice of bases are
\begin{equation}
  \label{eqAIJ}
  A^l_{ij} = \langle A( \phi_j ), e_i \rangle = \sum_{t \in \mathcal{T}} \mu( X_t ) \phi_j( X_t ) X_t^i, 
\end{equation}
where $ \langle \cdot, \cdot \rangle $ is the inner product of $ H $, and $ X_t^i = \langle X_t, e_i \rangle $. Computing the SVD of the $ n \times l $ matrix $ [ A_{il}^l ] $ then leads to a spectral decomposition of  $ A^l $ analogous to~\eqref{eqSVD}, viz. 
\begin{equation}
  \label{eqASVD}
  A^l_{ij} = \sum_{k=1}^r u_{ik} \sigma^l_k v_{jk}, \quad r =  \min\{ l, n  \},
\end{equation}
where $ u_{ik} $ and $ v_{jk} $ are elements of $ n \times n $ and $ l \times l $ orthogonal matrices, respectively, and $ \sigma^l_k \geq 0 $ are singular values (ordered in order of decreasing $ \sigma^l_k $). Each $ u_k = ( u_{1k}, \ldots u_{nk} ) $ is a spatial pattern in $ H $ expanded in the $ \{ e_i \} $ basis. Moreover, the entries $ ( v_{1k}, \ldots v_{v_lk} ) $ are the expansion coefficients of the corresponding  temporal pattern in the $ \{ \phi_i \} $ basis for $ V_l $, leading to the temporal process
\begin{equation}
  \label{eqVK}
  v_k( t ) = \sum_{i=1}^l v_{ik} \phi_{i-1}( X_t )
\end{equation}
Thus, the decomposition of the signal $ X_t $ in terms of the $ l $ chronos and topos modes of $ A^l $, analogous to~\eqref{eqXKTSSA}, is 
\begin{equation}
  \label{eqXNLSA}
  X_t^l = \sum_{k=1}^l \tilde X_t^k, \quad \text{with} \quad  \tilde X_t^k = u_k \sigma^l_k v_k( t ). 
\end{equation}
Note that by completeness of Laplace-Beltrami eigenfunctions [$ \sum_{k=1}^\infty \phi_k( X ) \phi_k( X' ) = \delta( X - X' ) $], $ X_t^l $ converges to $ X_t $ as $ l \to \infty $. Moreover, the spatial and temporal covariance operators in NLSA, $ A^lA^{l*} $ and $ A^{l*} A^l $, can be expressed as convolutions of the spatial and temporal two-point correlation functions with the spectral kernel, $ \mathcal{ K }_l( X, X' ) = \sum_{k=1}^l \phi_k( X ) \phi_k( X' ) $,  weighted by the Riemannian measure; see \citep[][]{GiannakisMajda12a} for details. 

To complete the procedure, in step~(5) the $ \tilde X^k_t $ are projected to $ d $-dimensional physical space by writing 
\begin{equation}
  \tilde X^k_t = 
  \begin{bmatrix}
    \hat x_{t,0} \\
    \hat x_{t,\delta_t} \\
    \vdots \\
    \hat x_{t,(q-1)\,\delta_t}
  \end{bmatrix},
\end{equation}
 and taking the average, 
\begin{equation}
  \tilde x^k_t = \sum_{t',\tau:  t'-\tau = t} \hat x_{t',\tau} / q. 
\end{equation}
This leads to the decomposition in~\eqref{eqXTK}.
 
\subsection{Graph-theoretic analysis}

In applications, the Laplace-Beltrami eigenfunctions for a finite dataset are computed by replacing the manifold by a weighted graph $ G $ with vertices  $  \{ X_{\delta t }, X_{2 \, \delta t }, \ldots, X_{s \, \delta t} \}  \subset M $, and solving the eigenproblem of a transition probability matrix $ P $ defined on the vertex set of $ G $,  such that, for large-enough $ s $ and small-enough $ i $, the right eigenvectors $ \underline{ \phi }_i $ of $ P $ approximate the corresponding Laplace-Beltrami eigenfunctions $ \phi_i $ in~\eqref{eqLapl}; i.e., 
\begin{equation}
  \label{eqGraphEigenvalueProblem}
  P \underline{ \phi }_i = ( 1 - \lambda_i ) \underline{ \phi }_i, 
\end{equation}
with $ \underline \phi_i = ( \phi_{i1}, \ldots, \phi_{is} )^T $and $ \phi_{ij} \approx \phi_i( X_{ (j-1) \, \delta t } ) $. These eigenvectors satisfy an orthonormality condition which is the discrete analog to~\eqref{eqLapl}, 
\begin{equation}
  \sum_{k=1}^s \mu_k \phi_{ik} \phi_{jk} = \delta_{ij}, 
\end{equation}
with $ \mu_k $ given by the invariant measure (leading left eigenvector) of $ P $; namely,\begin{equation} 
  \vec{ \mu } = \vec{ \mu } P,
\end{equation}
where $ \vec{ \mu } = ( \mu_1, \ldots, \mu_s ) $, $ \mu_i > 0 $, and $ \sum_{\mu=1}^s \mu_i = 1 $.

In the present work, we evaluate $ P $ using the diffusion map (DM) algorithm of Coifman and Lafon \cite{CoifmanLafon06}, with a simple but important modification in the calculation of the Gaussian weight matrix. Specifically, we assign to each sample $ X_{i\delta t } $ a local velocity in embedding space, $ \xi_i = \lVert X_{i\, \delta t} - X_{(i-1)\, \delta t } \rVert $, and evaluate the Gaussian weights 
\begin{equation}
  \label{eqW}
  W_{ij} = \exp( - \lVert X_{i\, \delta t } - X_{j\, \delta t} \rVert ^2 / \epsilon ( \xi_i \xi_j )^{1/2} ), 
\end{equation}
where $ \lVert \cdot \rVert $ denotes the norm of $ H $. This approach was motivated by the clustering algorithm developed in \cite{ZelnikManorPerona04}, with the difference that in the latter paper $ \epsilon_i $ is evaluated using spatial nearest neighbors, rather than the temporal nearest neighbors employed here. Locally-scaled kernels have also been used in other works in the literature; e.g., \cite{SingerEtAl09,RohrdanzEtAl11}. 

In the standard implementation of DM, $ \epsilon $ must be sufficiently small in order for the diffusion process represented by $ P $ to be sensitive only to local neighborhoods around each data point. Here, the normalization by $ \xi_i $ enforces geometrical localization even for $ \epsilon = O( 1 ) $. In \citep[][]{GiannakisMajda12a}, we found that this type of adaptive kernel significantly enhances the capability of NLSA to capture rare events. The remaining standard steps needed to compute $ P $ given $ W $ are \citep[][]{CoifmanLafon06}
\begin{subequations}
  \begin{align}
    Q_i &= \sum_{j=1}^s W_{ij}, \\
    K_{ij} &= W_{ij} / ( Q_i Q_j ), \\
    D_i &= \sum_{j=1}^s K_{ij} \\
    P_{ij} &= K_{ij} / D_i.
  \end{align}
\end{subequations}
Because the Riemannian measure in the DM family of algorithms is given by 
\begin{equation}
  \label{eqMu}\mu_i \propto \sum_{j=1}^{s} \frac{ W_{ij} }{ \left( \sum_{k=1}^{s} W_{ik} \right)^\alpha \left( \sum_{r=1}^{s} W_{jr} \right)^\alpha}
\end{equation}
for some parameter $ \alpha \in \mathbb{ R } $, states with large $\xi_i$ acquire larger Riemannian measure than in the uniform-$ \xi $ case. Hereafter, we work with the so-called Laplace-Beltrami normalization, $ \alpha = 1 $  \cite{CoifmanLafon06}. Even though we have not performed such tests, we believe that suitable basis functions for NLSA may be constructed using a variety of other graph Laplacian algorithms (e.g., \citep{BelkinNiyogi03,LeeVerleysen07}), so long as those algorithms permit the use of time-adaptive weights of the form of~\eqref{eqW}.  

The scalability of this class of algorithms to large problem sizes has been widely demonstrated in the machine learning and data mining literature. In particular, as a result of Gaussian decay of the weights, the $ W $ matrix used in implementations is made sparse, e.g., by truncating $ W $ to the largest $ b $ nonzero elements per row with $ b / s \ll 1 $, significantly reducing the cost of the eigenvalue problem for $ \underline{ \phi }_i $. The least-favorable scaling involves the pairwise distance calculation between the data samples in embedding space, which scales like $ s^2 \dim( H ) $ if done in brute force. Despite the quadratic scaling with $ s $, the linear scaling with $ \dim( H ) $ is of key importance, as it guarantees that NLSA does not suffer from a ``curse of dimension'' as do neural-network-based methods \cite{Hsieh07}. Moreover, an  $ s \log s $ scaling may be realized in the pairwise-distance calculation if the dimension of $ H $ is small-enough for approximate $ kd $-tree-based algorithms to operate efficiently \cite{AryaEtAl98}. Note also that the number of elements of the $ n \times l $ matrix $ A^l $ \eqref{eqASVD} appearing in the SVD step of NLSA is smaller by a factor of $ s / l $ compared to the full $ n \times s $ data matrix $ X $ \eqref{eqX} used in SSA, resulting in significant efficiency gains in practical applications with $ s \gg l $.  In the present study, all calculations were performed on a desktop workstation using brute-force evaluation of pairwise distances. 

 \subsection{\label{secDimensionSelection}Selecting the dimension of $ V_l $ via spectral entropy}

An important question concerning parameter selection in NLSA is how to choose the parameter $ l $ controlling the dimension of the temporal spaces $ V_l $. Clearly, working at large  $ l $ is desirable, since the lengthscale on the data manifold resolved by the eigenfunctions spanning $ V_l$ generally becomes smaller as $ l $ grows. However, for a given finite number of samples $ s $, the approximation error in the eigenvectors of the graph Laplacian in~\eqref{eqGraphEigenvalueProblem} also  increases with $ l $ \citep[][]{CoifmanLafon06}. In other words, the eigenfunctions $ \phi_i $ determined through graph-theoretic analysis will generally depend more strongly on $ s $  for large $ i $, resulting in an overfit of the discrete data manifold. Thus, in practical applications it is important to establish criteria that allow one to determine a reasonable tradeoff between improved resolution and risk of overfitting. 

One way of addressing this issue is to monitor the growth of an operator norm for $ A_l $ with $ l $, seeking plateau behavior or $ L $-shaped curves. A standard choice of operator norm in this context is the Frobenius norm, which may be evaluated conveniently via the matrix elements $ A_{ij}^l $ in~\eqref{eqAIJ}, viz.
\begin{equation}
  \label{eqFrobeniusNorm}
  \lVert A^l \rVert^2 = \sum_{i=1}^n \sum_{j=0}^l ( A_{ij}^l )^2.
\end{equation}
However, as we will see below, this approach may lead to considerable uncertainty in the choice of $ l $. Instead, we find that a more effective method is to choose $ l $ by monitoring changes in the distribution of the singular values $ \sigma^l_i $ with $ l $. In particular, we propose to assess the behavior of the spectrum of $ A^l $ with $ l $ via a spectral entropy measure, as follows. 

Let 
\begin{equation}
  p^l_i = \frac{ ( \sigma^{l}_i )^2}{ \sum_{j=1}^l (\sigma_j^l)^2 } 
\end{equation}
be normalized weights measuring the distribution of energy among the spatiotemporal modes of $ A^l $, which, as usual, are ordered in order of decreasing $ \sigma_i^l $. Consider also the energy distribution $ \pi^{l+1}_i $ over $ l + 1 $ modes determined by replicating $ \sigma^l_i $ for $ i \in [ 1, l ] $, and setting the energy in $ \pi^{l+1}_{l+1} $ equal to $ ( \sigma^l_l )^2 $. That is, we have
\begin{equation}
  \pi^{l+1}_i = \frac{ \hat\sigma^2_i }{ \sum_{j=1}^{l+1} \hat\sigma^2_j }, \quad \text{with} \quad 
  \hat\sigma_i = 
  \begin{cases}
    \sigma^l_i, \quad i \leq l, \\
    \sigma^l_{i-1}, \quad i = l + 1.
  \end{cases}
\end{equation}
Here, we measure the change in the spectrum of $ A^l $ relative to $ A^{l+1} $ through the relative entropy between the energy distributions $ p^{l+1}_i $ and $ \pi^{l+1}_i $, i.e.,
\begin{equation}
  \label{eqDL}
  D_l = \sum_{i=1}^{l+1} p^{l+1}_i \log \frac{ p^{l+1}_i }{ \pi^{l+1}_i }.
\end{equation}
It is a standard result in information theory and statistics \citep[][]{CoverThomas06} that $ D_l $ is a non-negative quantity which vanishes if and only if $ p^{l+1}_i = \pi^{l+1}_i $ for all $ i $. In Sec.~\ref{secDimensionSelection2} we demonstrate that as $ l $ grows $ D_l $ exhibits a sequence of spikes at small to moderate $ l $ (as qualitatively new features appear in the spectrum of $ A^l $), until it eventually settles to small values at higher $ l $. The practical criterion proposed here is to set $ l $ to values near that transition. 

\section{\label{secModes}Spatiotemporal patterns in the North Pacific sector of a comprehensive climate model}

\subsection{Dataset description}

We apply the NLSA framework presented above to study variability in the North Pacific sector of CCSM3; specifically, variability of the mean upper 300 m sea temperature field in the 700 y equilibrated control integration used by Teng and Branstator \cite{TengBranstator10} and Branstator and Teng \cite{BranstatorTeng10} in work on the initial and boundary-value predictability of subsurface temperature in that model. Here, our objective is to diagnose the prominent modes of variability in a time series generated by a coupled general circulation model. In this analysis, the $ x_t $ observable is the mean upper 300 m temperature field sampled every month at $ d = 534 $ gridpoints (native ocean grid mapped to the model's T42 atmosphere) in the region 20$^\circ$N--65$^\circ$N and 120$^\circ$E--110$^\circ$W.

\subsection{\label{secSpatiotemporalPatterns}Spatiotemporal patterns revealed by NLSA}

Deferring a discussion on parameter selection to Sec.~\ref{secDiscussion}, we begin with a description of the spatiotemporal patterns determined via NLSA using a two-year lagged-embedding window and the leading $ 27 $ Laplace-Beltrami eigenfunctions as basis functions for $ V_l $. Thus, the dimension of the spatial and temporal spaces is $ n = d \times 24 = \text{12,816} $ and $ \dim( V_l ) = l = 27 $, respectively.  Throughout, we work with canonical Euclidean distances between vectors in embedding space, 
\begin{equation}
  \lVert X_{t} - X_{t'} \rVert ^2 = \sum_{i=1}^n ( X_{t}^i - X_{t'}^i )^2,
\end{equation}
where $ X_t^i = \langle e_i, X_t \rangle $ denotes the $ i $-th gridpoint value of the system state at time $ t $ in embedding space. For the evaluation of the graph-Laplacian eigenfunctions in~\eqref{eqGraphEigenvalueProblem} we nominally set $ \epsilon = 2 $ (adaptive kernel width) and $ b = 3500 $ (number of edges per data point). 

\begin{figure}
  \centering\includegraphics{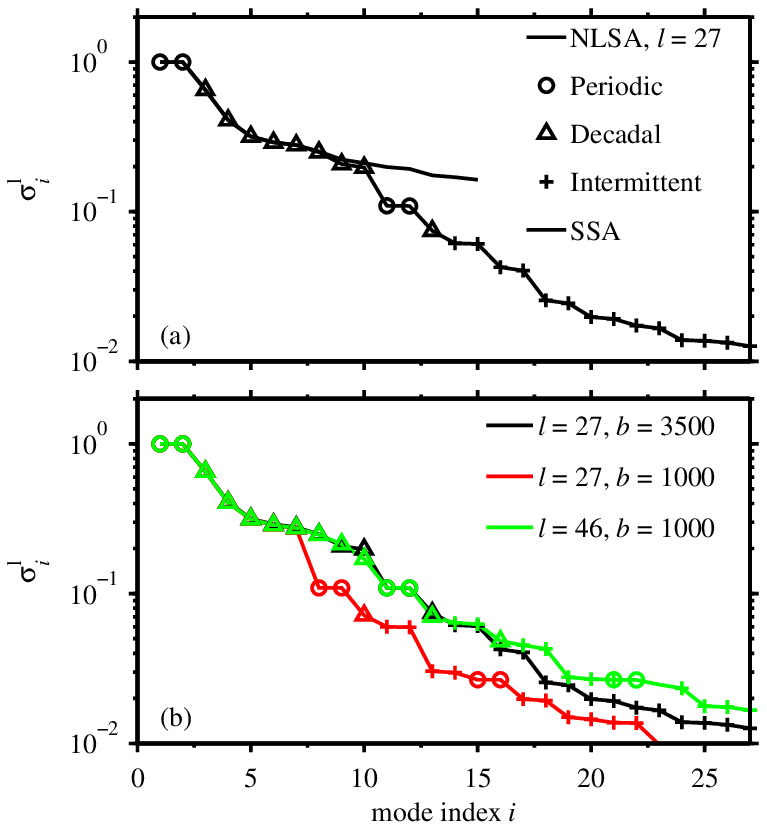}
  \caption{\label{figSingularValues}Singular values $ \sigma^l_i $ (normalized so that $ \sigma^l_1 = 1 $) for the periodic, low-frequency (decadal), and intermittent spatiotemporal patterns evaluated through NLSA and SSA for lagged embedding window $ \Delta t = 2 $ y. (a) NLSA spectrum for the nominal parameter values (temporal space dimension $l=27$, nearest neighbors per data point $b=3500$) and the corresponding spectrum from SSA. (b) Comparison of NLSA spectra with different $ l $ and $ b $ values.The kernel scaling parameter is $ \epsilon = 2 $ throughout.}
\end{figure}

 We display the singular values $ \sigma_i^l $ of the resulting $ A^l $ operator in Fig.~\ref{figSingularValues}(a). To verify the robustness of our results, we evaluated the spectrum of $ A^l $ for  various parameter values in the intervals $ \epsilon \in [ 1, 2 ] $, $ b \in [ 30, 4000 ] $, and $ l \in [ 10, 100 ] $. In the examples shown in Fig.~\ref{figSingularValues}(b), reducing $ b $ to 1000 with $ \epsilon $ and $ l $ fixed to their nominal values causes some reshuffling of the modes (e.g., $ \sigma_9^{27}$ and $ \sigma_{10}^{27} $), but imparts little change to the singular values of the modes which are not reshuffled. Increasing $ l $ to 47 reestablishes the mode ordering to some extent, bringing the $ b = 1000 $ and $ b = 3500 $ spectra to a near overlap for the first 17 modes.  We will return to the question of setting $ l $ in Sec.~\ref{secDimensionSelection2} ahead, but for the time being note that the modes which are most sensitive to changes of $ b $ and/or $ l $ in Fig.~\ref{figSingularValues} lie on the SSA branch of the spectrum, i.e., they are modes carrying little spectral information in the sense of the relative entropy metric $ D_l $ in~\eqref{eqDL}. In general, the properties the leading modes in the all three families described below exhibit small sensitivity to changes of $ \epsilon $ and $ b $ in the examined ranges, with $ l $ selected according to the $ D_l $ criterion. 

\begin{figure}
  \centering\includegraphics{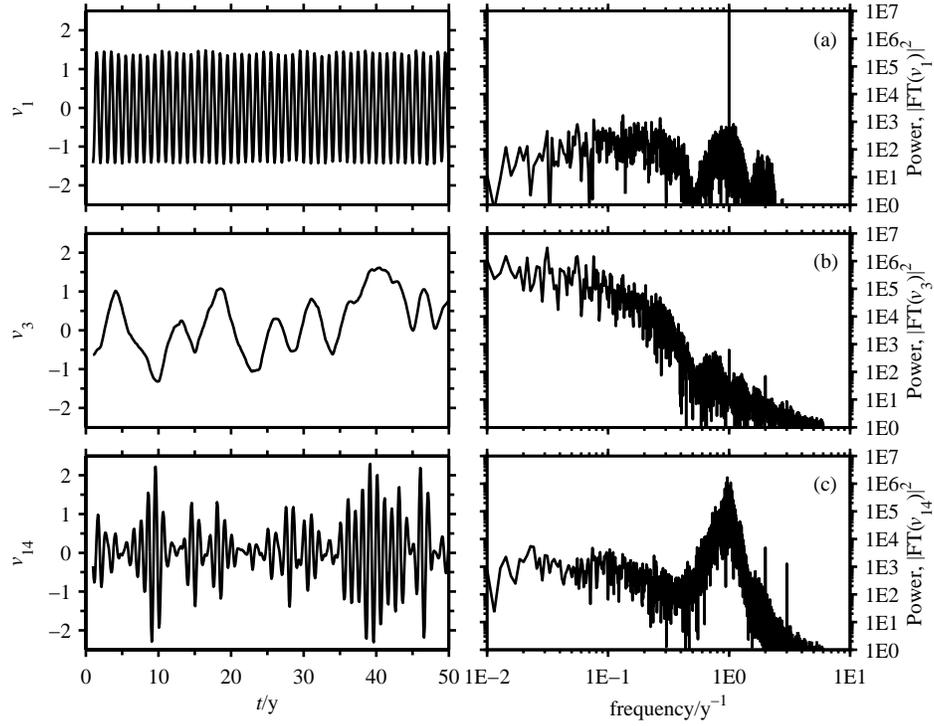}
  \caption{\label{figTemporalPatterns}Temporal patterns $ v_k( t ) $ corresponding to the singular values in Fig.~\ref{figSingularValues}. Shown in the temporal (left-hand panels) and frequency domains (right-hand panels) are (a) the annual mode, $v_1 $, (b) the PDO mode, $ v_3 $, (c) the leading Kuroshio intermittent mode, $ v_{14} $.}
\end{figure}

\begin{figure}
  \centering\includegraphics{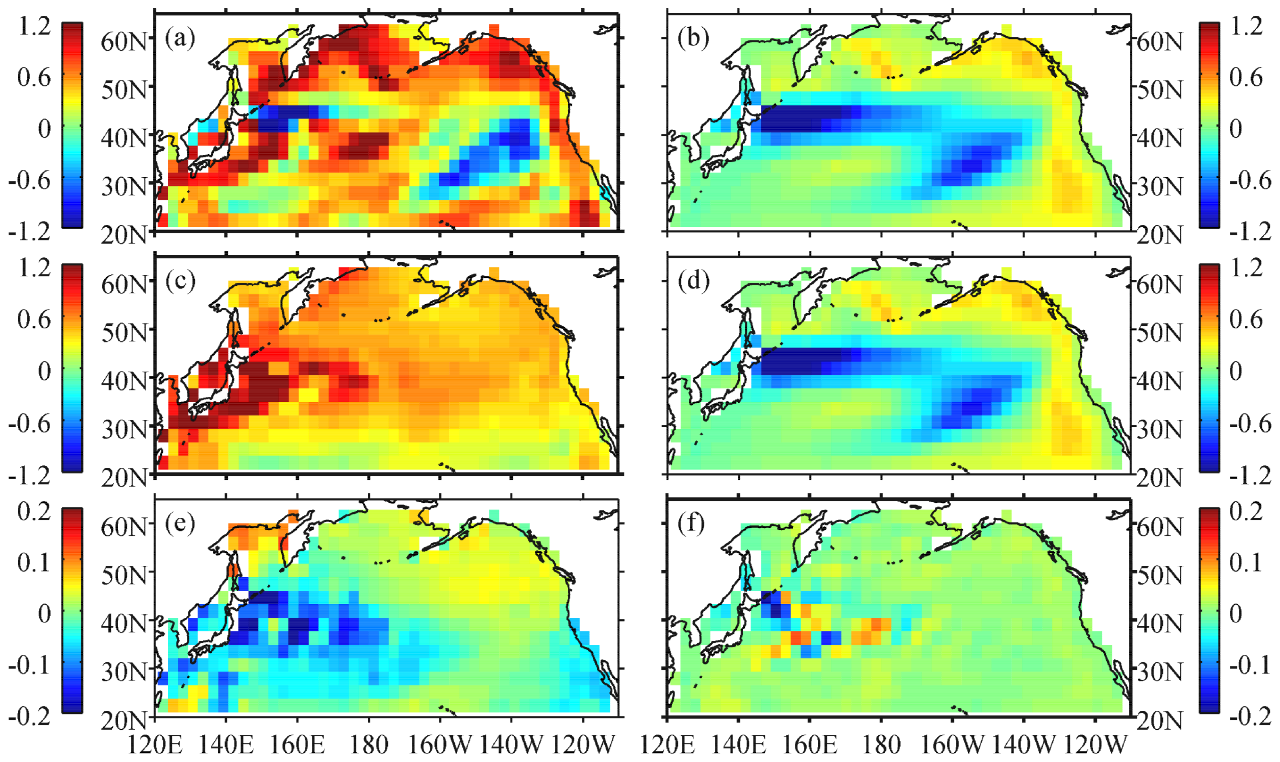}
  \caption{\label{figReconstruction}Spatiotemporal patterns of the the upper 300 m temperature anomaly field (annual mean subtracted at each gridpoint) in the North Pacific sector of CCSM 3, evaluated using NLSA with $ l =27 $ (see Fig.~\ref{figSingularValues}) and SSA. (a) Raw data in November of year 91 of Figure~\ref{figTemporalPatterns}. (b) Leading low-frequency mode from SSA. (c--f) Composite fields determined through NLSA. (c) Annual modes, $ v_1 $ and $ v_2 $. (d) Leading low-frequency mode, $ v_3 $, describing the PDO. (e) Semiannual modes, $ v_{11} $ and $ v_{12} $. (f) First two-fold degenerate set of intermittent modes, $ v_{14} $ and $ v_{15} $, describing variability of the Kuroshio current. The dynamical evolution of these patterns in Movie~S1 is much more revealing.}
\end{figure}

For the remainder of Sec.~\ref{secModes}, we work with the nominal values $ \epsilon = 2 $, $ b = 3500 $, and $ l = 27 $.  Studying the corresponding temporal patterns $ v_k $ from~\eqref{eqVK} in both the temporal and frequency (Fourier) domains, we find that the modes fall into three distinct families of periodic, low-frequency, and intermittent modes, illustrated in Fig.~\ref{figTemporalPatterns}, and described below. The resulting spatiotemporal patterns $ \tilde x^k_t $ from~\eqref{eqXTK} are shown in Fig.~\ref{figReconstruction} and, more clearly, in the dynamical evolution in Movie~S1. Note that the time-lagged embedding in~\eqref{eqEmbedding} is essential to the separability of the modes into these families; we will return to this point in Sec.~\ref{secLaggedEmbedding}.   

\textbf{Periodic modes.} The periodic modes come in doubly-degenerate pairs (see Fig.~\ref{figSingularValues}), and have the structure of sinusoidal waves with phase difference $ \pi / 2 $ and frequency equal to integer multiples of 1 y$^{-1} $. The leading periodic modes, $ v_1 $ and $ v_2 $,  represent the annual (seasonal) cycle in the data. In the physical domain [Fig.~\ref{figReconstruction}(c) and Movie~S1(c)],  these modes generate an annual oscillation of temperature anomalies, whose amplitude is largest ($\sim 1^\circ$C) in the western part of the basin ($\sim 130^\circ$E--$160^\circ$E) and for latitudes in the range $30^\circ$N--$45^\circ$N. The second set of periodic modes, $ v_{11} $ and $ v_{12} $, have semiannual variability. These modes exhibit significant amplitude in the western part of the domain [Fig.~\ref{figReconstruction}(e) and Movie~S1(e)], but also along the West Coast of North America, which is consistent with semiannual variability of the upper ocean associated with the California current \cite{MendelssohnEtAl04}. Together with the higher-frequency overtones, the modes in this family are the standard eigenfunctions of the Laplacian on the circle, suggesting that the data manifold $ M $ has the geometry of a circle along one of its dimensions.

\textbf{Low-frequency modes.} The low-frequency modes are characterized by high spectral power over interannual to interdecadal timescales, and strongly suppressed power over annual or shorter time scales. As a result, these modes represent the low-frequency variability of the upper ocean, which has been well-studied in the North Pacific sector of CCSM3 \citep{AlexanderEtAl06,BranstatorTeng10,TengBranstator10}. The leading mode in this family [$ v_3 $; see Fig.~\ref{figTemporalPatterns}(b)], gives rise to a typical PDO pattern [Figure~\ref{figReconstruction}(c) and Movie~S1(c)], where the most prominent basin-scale structure is a horseshoe-like temperature anomaly pattern developing eastward along the Kuroshio current, together with an anomaly of the opposite sign along the west coast of North America. The higher modes in this family gradually develop smaller spatial features and spectral content over shorter time scales than $ v_3 $, but have no spectral peaks on annual or shorter timescales. 

\textbf{Intermittent modes.} As illustrated in Fig.~\ref{figReconstruction}(f) and Movie~S1(f), the key feature of modes in this family is temporal intermittency, arising out of oscillations at annual or higher frequency, which are modulated by relatively sharp envelopes with a temporal extent in the 2--10-year regime. Like their periodic counterparts, the intermittent modes form nearly degenerate pairs (see Fig.~\ref{figSingularValues}), and their base frequency of oscillation is an integer multiple of 1 year$^{-1} $. The resulting Fourier spectrum is dominated by a peak centered at at the base frequency, exhibiting some skewness towards lower frequencies.        

In the physical domain, these modes describe processes with relatively fine spatial structure, which are activated during the intermittent bursts, and become quiescent when the amplitude of the envelopes is small. The most  physically-recognizable aspect of these processes is enhanced transport along the Kuroshio current region, shown for the leading-two intermittent modes ($ v_{14} $ and $ v_{15} $) in Figure~\ref{figReconstruction}(d). This process features sustained eastward propagation of small-scale,  $ \sim 0.2 $ $^\circ $C  temperature anomalies during the intermittent bursts. The intermittent modes higher in the spectrum also encode rich spatiotemporal patterns, including retrograde (westward) propagating anomalies, and gyre-like patterns resembling the subpolar and subtropical gyres. These features are shown in Movie~S1(f), which displays a composite temperature anomaly field consisting of the leading four intermittent modes ($ v_{14}, \ldots, v_{17} $; see Fig.~\ref{figSingularValues}). 

\section{\label{secDiscussion}Discussion}

\subsection{Intermittent processes and relation to SSA}

The main result of this analysis, which highlights the importance of taking explicitly into account the nonlinear structure of complex high-dimensional datasets, is the existence of intermittent patterns of variability in the North Pacific sector of CCSM3, which are not accessible through SSA. This type of variability naturally emerges by restricting the temporal modes to lie in the low-dimensional subspaces $ V_l $ spanned by the leading Laplace-Beltrami eigenfunctions on the data manifold $ M $. The inner product of these vector spaces, weighted by the Riemannian measure $ \mu $ in~\eqref{eqInnerProdVL}, plays an important role in the skill of NLSA of capturing intermittency and rare events \citep[][]{GiannakisMajda12a}. Heuristically, this is because $ \mu( X_t ) $ weighs each state $ X_t $ by the volume it occupies in $ M $, which is expected to be large for rare and/or intermittent states.  

As shown in Figs.~\ref{figSingularValues} and~\ref{figReconstruction},  the spatiotemporal patterns determined through NLSA are in close agreement with SSA for the annual and low-frequency modes, but intermittent modes have no SSA counterparts. In particular, instead of the qualitatively-distinct families of processes described above, the SSA spectrum is characterized by a smooth decay involving modes of progressively higher spatiotemporal frequencies, but with no intermittent behavior analogous, e.g., to mode $ v_{14} $ in Fig.~\ref{figTemporalPatterns}. The $ \sigma^l_i $ values associated with the intermittent modes and, correspondingly, their contributed variance of temperature anomaly, is significantly smaller than the periodic or low-frequency modes. However, this is not to say the dynamical significance of these modes is negligible. In fact, intermittent events, carrying low variance, are widely prevalent features of complex dynamical systems \cite{AubryEtAl93,CrommelinMajda04,GiannakisMajda12a}. Being able to capture this intrinsically nonlinear behavior constitutes one of the major strengths of the NLSA methodology presented here. 

\subsection{\label{secDimensionSelection2}Selecting the temporal-space dimension through spectral entropy}

As discussed in Sec.~\ref{secDimensionSelection}, an important issue in NLSA is the selection of the dimension of the temporal space $ V_l $ through the number $ l $ of Laplace-Beltrami eigenfunctions used in~\eqref{eqPhiL}. Setting $ l $ too low will cause some important features to be lost, due to under-resolution on the data manifold. On the other hand, as $ l $ grows, eventually the algorithm will overfit the data if the number of samples $ s $ remains fixed. Here, we illustrate how the spectral entropy measure $ D_l $ introduced in~\eqref{eqDL} can be used to inform the selection of suitable values of $ l $. For ease of comparison, we normalize $ D_l $ to lie in the unit interval by applying the standard transformation \cite{Joe89} 
\begin{equation}
  \label{eqDNormalize}
  D_l \mapsto ( 1 - \exp( - 2 D_l ) )^{1/2}.
\end{equation}

\begin{figure}
  \centering\includegraphics{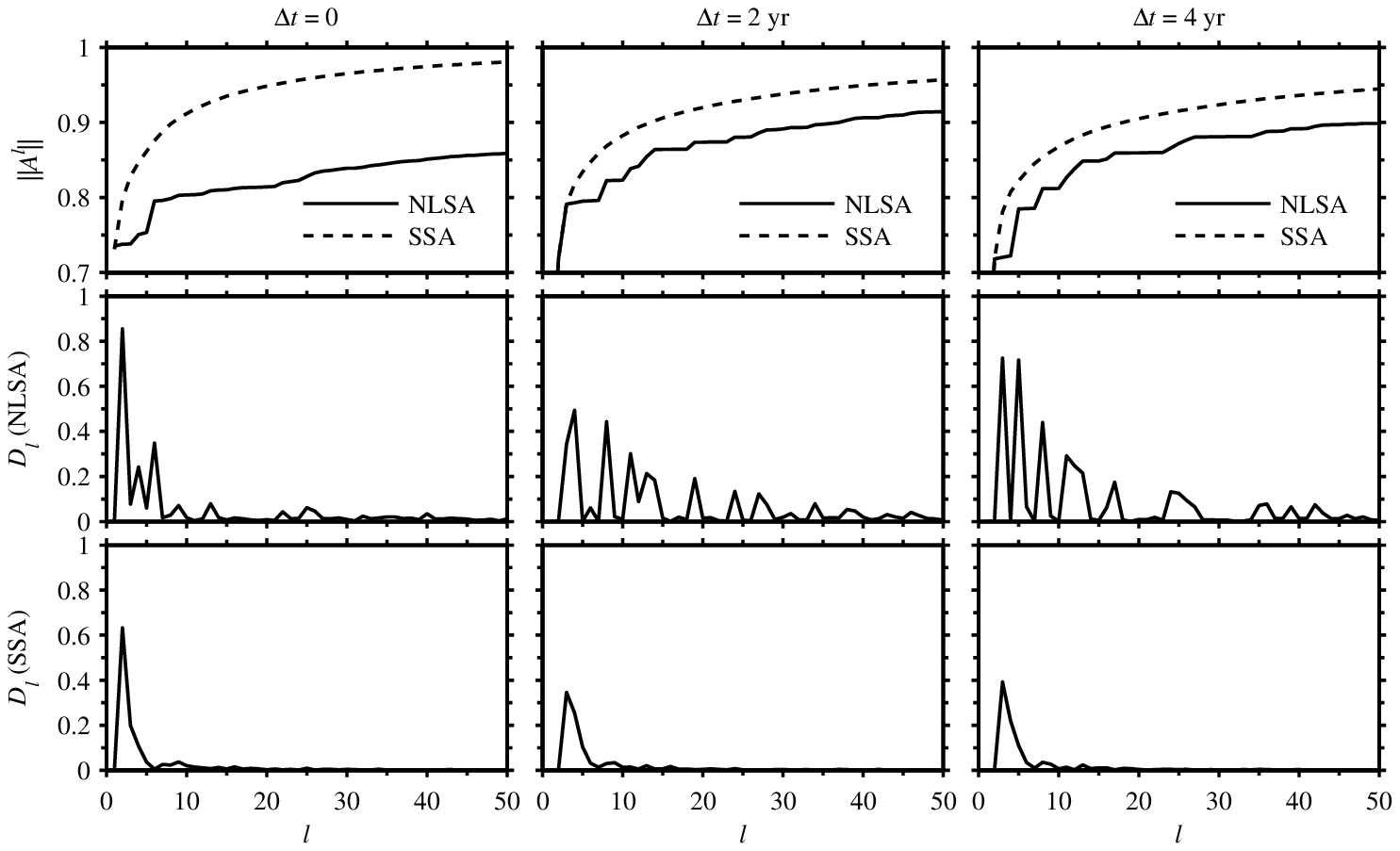}
  \caption{\label{figRelativeEntropy}Frobenius norm of the $ A^l $ operator and spectral entropy $ D_l $, evaluated via~\eqref{eqFrobeniusNorm} and~\eqref{eqDL}, versus the parameter $ l $ controlling the dimension of the temporal spaces $ V_l $ ($\dim(V_l)=l$). The lagged-embedding window is $ \Delta t = 0 $, 2, and 4 y. The norm $ \lVert A^l \rVert $ has been normalized to unity at $ l = s $ (number of samples). Also shown for reference are the corresponding Frobenius norms and spectral entropy measures evaluated by truncating the SSA expansion in~\eqref{eqXKTSSA} at $ l $ modes. Note that the Frobenius norms for NLSA and SSA cannot be compared directly because they are defined for linear maps acting on different vector spaces. $ D_l $ has been scaled to the unit interval by applying the transformation in~\eqref{eqDNormalize}.} 
\end{figure}

\begin{figure}
  \centering\includegraphics{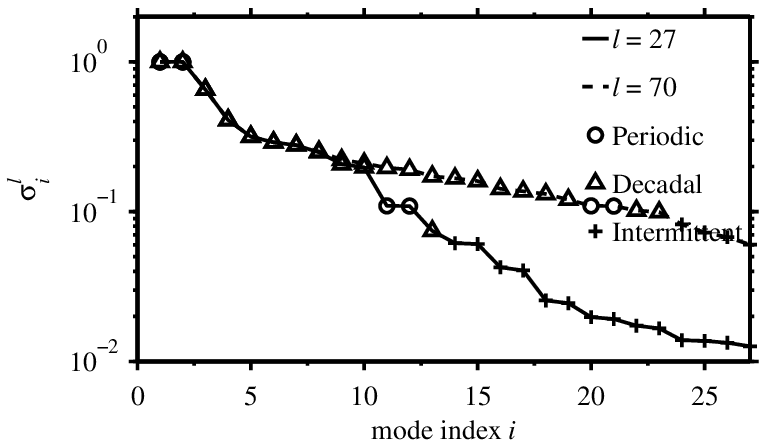}
  \caption{\label{figSingularValues2}Singular value spectrum from NLSA with temporal space dimension $ l = 23 $  and 70.}
\end{figure}

Figure~\ref{figRelativeEntropy} shows the dependence of $ D_l $, as well as the Frobenius norm $ \lVert A^l \rVert $ of the linear operators in NLSA, for values of $ l $ in the interval $ [ 1, 50 ] $ and embedding windows $ \Delta t = 0 $, 2, and 4 y. As expected, $ \lVert A^l \rVert $ increases with $ V_l $, and apart from the case with $ \Delta t =0 $, rapidly approaches values close to 90\% of its maximum value (occurring for $ l = s $). Compared with the corresponding behavior of the norm in truncated SSA expansions (also shown in Fig.~\ref{figRelativeEntropy}), $ \lVert A^l \rVert $ follows a more staircase-like growth, but because the operator norm is dominated by the leading few singular values carrying the majority of the energy of the signal, it is not immediately obvious when $ l $ has reached an optimum value. On the other hand, the spectral entropy measure $ D_l $ clearly transitions from a regime of saw-tooth behavior with appreciable magnitude at small to moderate $ l $, to a regime of negligible amplitude at larger values of $ l $.  That transition, which occurs around $ l = 25 $ for the NLSA cases with $ \Delta t = 2 $ and $ 4 $ y ($ l \simeq 7 $ for the case with no embedding and for SSA), indicates that increasing the dimension of $ V_l $ beyond those values introduces no qualitatively new spatiotemporal patterns. This provides justification for the $ l = 27 $ value used in Sec.~\ref{secModes}. As illustrated in Fig.~\ref{figSingularValues2}, the spectral gaps separating the low-frequency, semiannual, and intermittent modes are absent from the $ \sigma^l_i $ spectrum with $ l $ significantly exceeding the threshold value identified via $ D_l $.     
 
\subsection{\label{secLaggedEmbedding}The role of lagged embedding}

\begin{figure}
  \centering\includegraphics{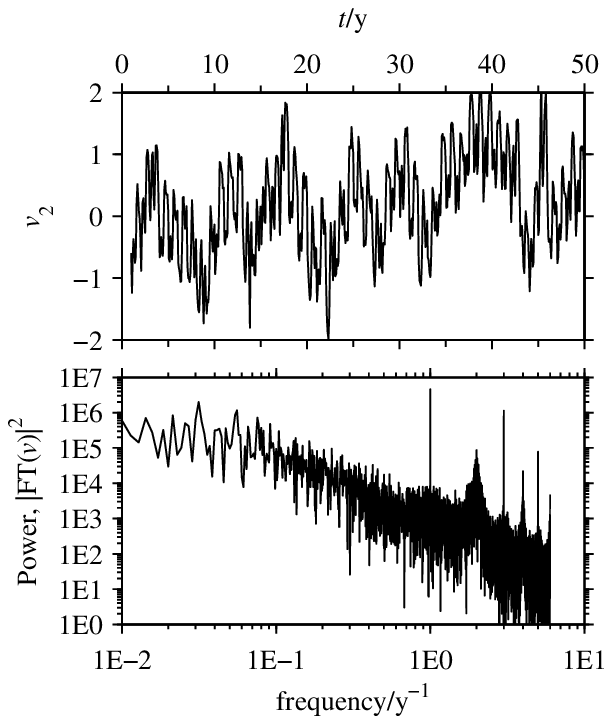}
  \caption{\label{figVTMixed}Leading ``low-frequency'' mode evaluated without embedding [cf.\ Fig.~(2b)]. Note the pronounced spectral lines with period $ \{ 1, 1/2, 1/3, \ldots, 1/6 \} $ y.}
\end{figure}

The embedding in~\eqref{eqEmbedding} of the input data $ x_t $ in $ H $ is essential to the separability of the temporal into distinct families of processes. To illustrate this, in Figure~\ref{figVTMixed} we display the temporal mode that most-closely resembles the PDO mode of Fig.~\ref{figReconstruction}(d), evaluated without embedding ($ q = 1 $, $ \Delta t = 0 $). It is evident from both the temporal and Fourier representations of that mode that the decadal process recovered in Sec.~\ref{secSpatiotemporalPatterns} using a two-year embedding window has been contaminated with high-frequency variability; in particular, prominent spectral lines at integer multiples of 1 y$^{-1} $ down to the maximum frequency of $6/\text{y} $ allowed by the monthly sampling of the data. An even stronger frequency mixing was found to take place in the corresponding temporal SSA modes. 

Optimal values for the embedding window are generally hard to determine \emph{a priori}. However, the following guidelines are useful in the parameter selection process if some physical information is available about the dataset:
\begin{enumerate}
\item \emph{External factors and non-Markovian effects.} If the data are known to be influenced by time-dependent external factors (e.g., solar forcing), or exhibit non-Markovian effects due to partial observations \cite{PackardEtAl80,Takens81,BroomheadKing86,SauerEtAl91}, then a good initial choice for $ \Delta t $ would be a value comparable to timescales associated with those factors. The $ \Delta t = 2 $ y choice made here is compatible with the annually varying solar forcing, which is one of the most important external factors in upper ocean dynamics. 
\item \emph{Timescales associated with propagating structures.} If the dataset features  propagating coherent structures (e.g., oceanic or atmospheric waves), then suitable candidate values for $ \Delta t $ are given by the natural timescales (e.g., period, decorrelation time) of the propagating structures. Because each point in lagged embedding space corresponds to a spatiotemporal process of temporal extent $ \Delta t $, such propagating structures can be described via two-fold degenerate sets of modes evolving in temporal quadrature. For instance, the $ \Delta t = 2 $ y window employed here is long-enough to capture prominent low-frequency waves in the ocean such as baroclinic Rossby waves and intraseasonal Kelvin waves \cite{CheltonSchlax96}. See \cite{GiannakisEtAl12} for an example involving organized convection in the tropics. 
\end{enumerate}
The relative entropy measure $ D_l $ of Sec.~\ref{secDimensionSelection} can also be used to inform the selection of $ \Delta t $, especially in situations with little \emph{a priori} information about the dynamics.  In separate calculations, we have verified that the modes separate into periodic, low-frequency, and intermittent processes for embedding windows up to $ \Delta t = 10 $ y, including the $ \Delta t = 4 $ y case displayed in Fig.~\ref{figRelativeEntropy}.

\subsection{\label{secDeseasonalizing}Seasonally-detrended data}

A standard preprocessing step in the geosciences and other disciplines is to detrend the data by subtracting monthly climatology, or applying bandpass filtering, to isolate the signal of interest (see, e.g., \cite{VonStorchZwiers02,WaliserEtAl09} for an overview of popular methods). In the context of NLSA algorithms,  we advocate that such preprocessing can in fact lead to significant loss of information and performance degradation. Conceptually, preprocessing a nonlinear data manifold through a linear operation such as seasonal-cycle subtraction may lead to distortion of the neighborhood relationships of the data, with adverse consequences on the extracted spatiotemporal patterns, which is what we have observed in the North Pacific data studied here. 
 
\begin{figure}
  \centering\includegraphics{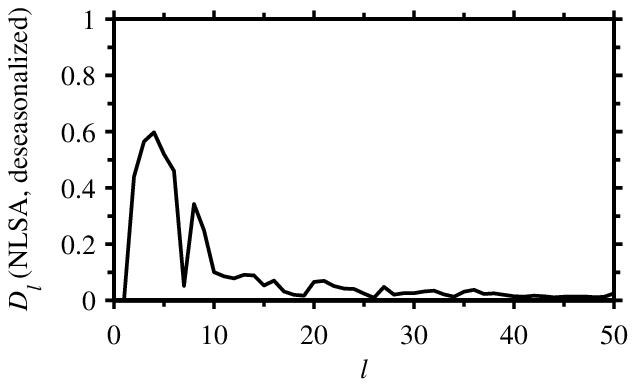}
  \caption{\label{figRelativeEntropyDeseasonalized}Relative entropy $ D_l $, normalized to the unit interval via~\eqref{eqDNormalize}, as a function of temporal space dimension $ l $ for deseasonalized data. The lagged embedding window is $ \Delta t = 2 $ y.}
\end{figure}

\begin{figure}
  \centering\includegraphics{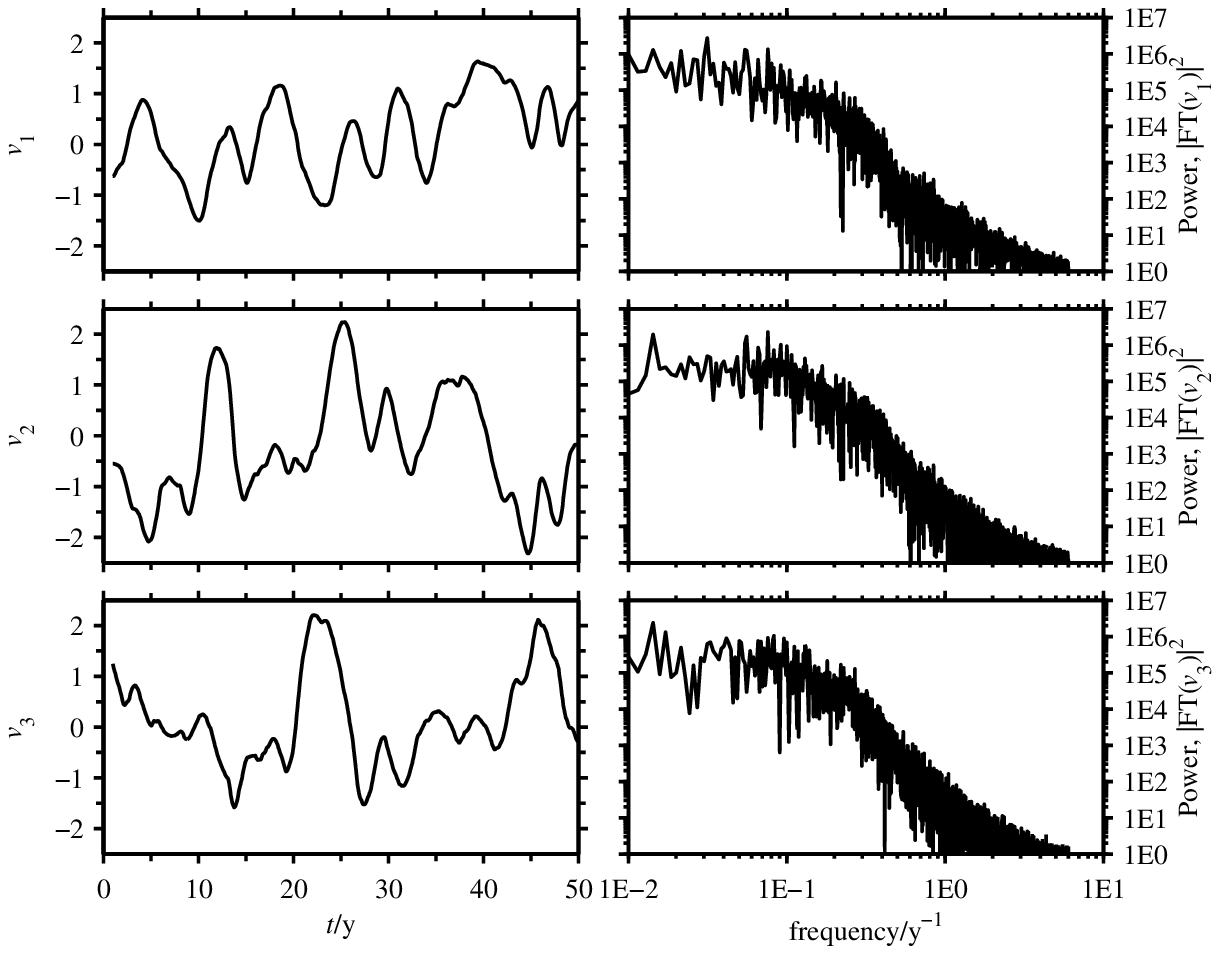}
  \caption{\label{figVTDeseasonalized}Leading three temporal patterns from NLSA for deseasonalized data. The lagged embedding window and temporal space dimension are $ \Delta t = $ 2 y and $ l = 11 $, respectively.}
\end{figure}

Figure~\ref{figRelativeEntropyDeseasonalized} shows the relative entropy measure $ D_l $ obtained by applying NLSA with lagged embedding window $ \Delta t = 2 $ y to the temperature field data of Sec.~\ref{secModes} with the monthly climatology (average temperature anomaly field for each month) subtracted. As manifested by the faster decay of $ D_l $ compared to the $ \Delta t = 2 $ y results of Fig.~\ref{figRelativeEntropy}, detrending leads to a decrease in the number of modes that carry significant information. Indeed, the temporal patterns of the detrended data all have the same qualitative structure as the low-frequency modes of Sec.~\ref{secSpatiotemporalPatterns}, characterized by decaying frequency spectra with no obvious peaks; see Fig.~\ref{figVTDeseasonalized}.  That seasonal detrending eliminates the annual periodic modes and their harmonics from the spectrum is intuitively obvious, but it turns out that the intermittent modes are removed from the spectrum as well. In particular, we have not been able to find modes with the spatial features characteristic of the intermittent modes in Fig.~\ref{figReconstruction}(e,f) in a number of experiments with temporal space dimension $ l $ up to 34. We believe that the intermittent modes depend crucially on phase relationships with the seasonal cycle and its higher harmonics, and as result the algorithm fails to detect these modes if seasonality has been removed from the data.  
  
\section{\label{secConclusions}Conclusions}

Combining techniques from machine learning and the qualitative theory of dynamical systems, in this work we have presented a scheme called NLSA for spatiotemporal analysis of high-dimensional time series, which takes explicitly into account the nonlinear geometrical structure of datasets arising in geophysics and other applied sciences. Like classical SSA \cite{GhilEtAl02}, the method presented here utilizes time-lagged embedding and SVD to produce a decomposition of time series generated by partial observations of high-dimensional, complex dynamical systems into distinct spatiotemporal modes. However, the linear operators used here in the SVD step differs crucially from SSA in that their domains of definition are low-dimensional Hilbert spaces of square-integrable functions on the nonlinear manifold $ M $ comprised by the data (in a suitable coarse-grained representation via a graph). This family of spaces, $ V_l $, is tailored to the nonlinear geometry of $ M $ through its Riemannian measure, and has high skill in capturing intermittency and rare events. As its dimension $ l $ grows, $ V_l $ provides a description of spatiotemporal patterns at increasingly fine resolution on the data manifold. Moreover, well-behaved orthonormal basis functions for these spaces can be computed efficiently via graph-Laplacian algorithms developed in data mining and machine learning \citep[][]{BelkinNiyogi03,CoifmanLafon06}. 

Applying this scheme to the upper-ocean temperature in the North Pacific sector of the CCSM3 model, we find a family of intermittent processes which are not captured by SSA. These processes describe eastward-propagating, small-scale temperature anomalies in the Kuroshio current region, as well as retrograde-propagating structures at high latitudes and in the subtropics (see Movie~S1). Moreover, they carry little variance of the raw signal, and display burst-like behavior characteristic of strongly nonlinear dynamics. The remaining identified modes include the familiar PDO pattern of low-frequency variability, as well as annual and semiannual periodic processes. The ability to recover the intermittent modes relies crucially on retaining the seasonal cycle (monthly climatology) of the data, suggesting that these modes operate in non-trivial phase relationships with the periodic modes.

The nature of the analysis presented here is purely diagnostic. In particular, we have not touched upon the dynamical role of these modes in reproducing the observed dynamics, e.g., by triggering large-scale regime transitions \citep[][]{CrommelinMajda04,OverlandEtAl08}. This question was addressed to some extent in \citep[][]{GiannakisMajda12a} in the context of a low-order chaotic model for the atmosphere, but remains open in applications involving high-dimensional complex dynamical systems with unknown equations of motion. Here, statistical modeling techniques, such as Bayesian hierarchical modeling \citep[][]{WikleBerliner07}, combined with information-theoretic methods for assessing predictability and model error \citep[][]{GiannakisMajda12b,GiannakisMajda12c}, are promising methods for training empirical models for these processes, and assessing their predictive skill. We plan to study these topics in future work. 

\section*{Acknowledgments}
This work was supported by NSF grant DMS-0456713, ONR DRI grants N25-74200-F6607 and N00014-10-1-0554, and DARPA grants N00014-07-10750 and N00014-08-1-1080.  We thank G.\ Branstator and H.\ Teng for providing access to the CCSM3 dataset used in this analysis.

\section*{Supporting information available}

Movie S1. Spatiotemporal patterns of the the upper 300 m temperature anomaly field (annual mean subtracted at each gridpoint) in the North Pacific sector of CCSM 3, evaluated using NLSA with $ l =27 $ (see Fig.~\ref{figSingularValues}) and SSA. (a) Raw data. (b) Leading low-frequency mode from SSA. (c--f) Composite fields determined through NLSA. (c) Annual modes, $ v_1 $ and $ v_2 $. (d) Leading low-frequency mode, $ v_3 $, describing the PDO. (e) Semiannual modes, $ v_{11} $ and $ v_{12} $. (f) Leading four intermittent modes, $ v_{14}, \ldots v_{17} $, describing variability of the Kuroshio current and retrograde (westward) propagating structures. The starting time of this animation is the same as in Fig.~\ref{figTemporalPatterns}.

\end{document}